\documentclass[aps,prl,twocolumn,superscriptaddress,showpacs]{revtex4}

\usepackage{graphicx}

\begin{document}

\title{Evidence for charge Kondo effect in superconducting Tl-doped PbTe}


\author{Y. Matsushita}
\affiliation{Department of Materials Science and Engineering and Geballe Laboratory for Advanced Materials}
\author{H. Bluhm}
\affiliation{Department of Physics and Geballe Laboratory for
Advanced Materials}
\author{T. H. Geballe}
\author{I. R. Fisher}
\affiliation{Department of Applied Physics and Geballe Laboratory for
Advanced Materials, Stanford University, Stanford, California
94305-4045}

\date{\today}

\begin{abstract}
We report results of low-temperature thermodynamic and transport
measurements of Pb$_{1-x}$Tl$_x$Te single crystals for Tl
concentrations up to the solubility limit of approximately $x =
1.5\%$. For all doped samples, we observe a low-temperature
resistivity upturn that scales in magnitude with the Tl
concentration. The temperature and field dependence of this upturn
are consistent with a charge Kondo effect involving degenerate Tl
valence states differing by two electrons, with a characteristic
Kondo temperature $T_K \sim 6$ K. The observation of such an effect
supports an electronic pairing mechanism for superconductivity in
this material and may account for the anomalously high $T_c$ values.
\end{abstract}


\maketitle

The Kondo effect arises from the interaction of conduction electrons
with degenerate degrees of freedom in a material and is usually
associated with dilute magnetic impurities in a nonmagnetic host. In
such cases, the two degenerate states correspond to the impurity
spins oriented up or down. Second order scattering processes
involving virtual intermediate states lead to the well-known
logarithmic increase in resistivity at low temperatures, which
saturates in the unitary scattering limit below a characteristic
Kondo temperature \cite{Hewson}. However, other systems comprising
two degenerate degrees of freedom can also lead to Kondo-like
phenomena \cite{Cox_1998}. In particular, a ``charge Kondo effect,''
corresponding to dilute impurities with two degenerate charge
states, has been proposed in the negative-$U$ Anderson model
\cite{Taraphder_1991}, though to date there has not been an
experimental realization of such an effect. Significantly, the
quantum valence fluctuations implicit in such a model, which involve
pairs of electrons that tunnel on and off impurity sites, also
provide an electronic pairing mechanism for superconductivity
\cite{Dzero_2004, Schuttler_1989, Hirsch_1985, Malshukov_1991}.

Thallium is one of several elements that is known to skip valences,
such that only Tl$^{1+}$ and Tl$^{3+}$ are observed in ionic
compounds, corresponding to electron configurations $6s^2$ and
$6s^0$ respectively. Compounds that one would otherwise expect to
contain divalent Tl are found to disproportionate. For example,
TlBr$_2$ is more specifically
Tl$^{\text{I}}$Tl$^{\text{III}}$Br$_{4}$, and TlS is likewise
Tl$^{\text{I}}$Tl$^{\text{III}}$S$_2$ \cite{Robin_1967}. This effect
is driven by the stability of a filled shell in conjunction with the
polarizability of the material. In this case, Tl$^{2+}$ can be
characterized by a negative effective $U$, where $U_n =
(E_{n+1}-E_n) - (E_n-E_{n-1}) < 0$ and $n$ labels the valence state
\cite{Varma_1988, Drabkin_1981, Moizhes_1982}. For this reason,
valence-skipping elements provide experimental access to
negative-$U$ behavior and are therefore suitable candidate
impurities for realizing a charge Kondo effect in a bulk material.

In this letter we describe measurements of PbTe doped with small
amounts of Tl. The material exhibits a Kondo-like upturn in the
resistivity in the absence of magnetic impurities and superconducts
at low temperatures. We make the case that this behavior originates
from a charge Kondo effect involving degenerate valence states of
the Tl impurities. This is the first evidence for such an effect and
is especially significant in substantiating claims that the
superconducting pairing mechanism in Tl-doped PbTe derives
principally from negative-$U$ effects.

PbTe is a small gap semiconductor. It has a rocksalt structure and
has been treated with reasonable success using ionic models (i.e.,
Pb$^{2+}$Te$^{2-}$) \cite{Weiser_1981}. The material can be doped to
degeneracy by either vacancies or third-element dopants, with
typical carrier concentrations in the range of 10$^{18} - 10^{20}$
cm$^{-3}$ \cite{Ravich, Nimtz, Khokhlov}. In comparison with similar
semiconducting materials such as SnTe, GeTe, and InTe, it was
previously anticipated that doped PbTe would only superconduct below
approximately 0.01 K, if at all \cite{Hulm_1970}. This has been
found to be the case for all dopants except thallium, for which
superconductivity was observed with critical temperatures up to 1.5
K \cite{Nemov_1998}, two orders of magnitude higher than anticipated
given the modest carrier concentrations. Given the anomalously high
$T_c$ values for Tl-doped PbTe, there has been considerable
discussion as to the role of the Tl impurities in this material,
whether they act as negative-$U$ centers, and specifically whether
such impurities can enhance an incipient tendency towards
superconductivity \cite{Moizhes_1983, Schuttler_1989, Hirsch_1985,
Krasinkova_1991, Dzero_2004}.

Single crystals of Pb$_{1-x}$Tl$_x$Te were grown by an unseeded
physical vapor transport method as described elsewhere \cite{Yana}.
The thallium content was measured by Electron Microprobe Analysis
(EMPA).  Resistivity measurements were made at 77 Hz and with
current densities in the range of 25 mA/cm$^2$ (corresponding to a
current of 10 $\mu$A for low-temperature measurements) to 1 A/cm$^2$
at higher temperatures. To check for heating effects, resistivity
data were taken for different current densities and for warming and
cooling cycles for each sample. Several samples were measured for
each Tl concentration.

\begin{figure}
\includegraphics{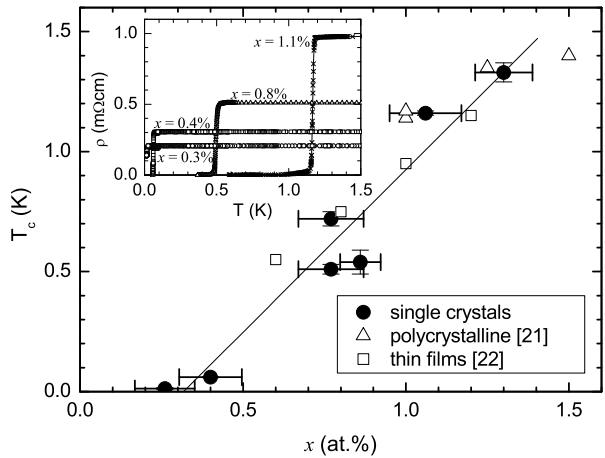}
\caption{\label{fig1}Variation of $T_c$ with Tl content $x$ for
Pb$_{1-x}$Tl$_x$Te. Line shows linear fit.  Inset shows
representative resistivity data for single crystals.}
\end{figure}

$T_c$ values estimated from the midpoints of superconducting
transitions in the resistivity are shown as a function of Tl
concentration in Fig.~\ref{fig1} and agree with published data for
polycrystalline samples \cite{Chernik_1981} and for thin films
\cite{Murakami_1996b} for high Tl concentrations. By measuring $T_c$
for the lowest Tl concentrations, we find that there is a critical
concentration of $\sim 0.3\%$ below which the material does not
superconduct above 20 mK. The inset to Fig.~\ref{fig1} shows the
sharp resistive transitions for representative samples.

\begin{figure}
\includegraphics{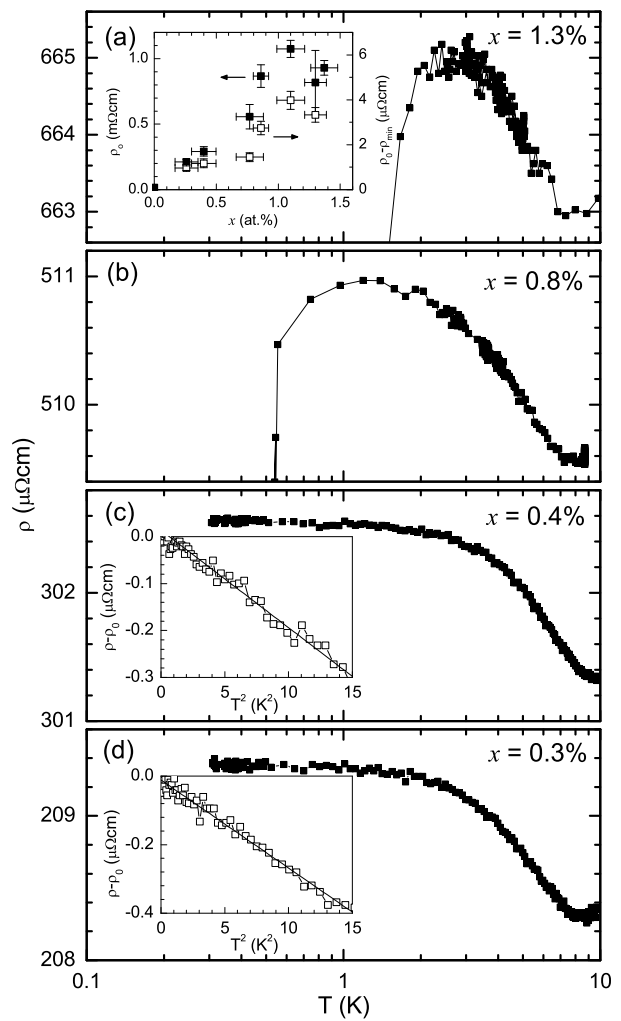}
\caption{\label{fig2}Low temperature resistivity of
Pb$_{1-x}$Tl$_x$Te for $x$ = 0.3, 0.4, 0.8, and 1.3$\%$. Inset to
panel (a) shows residual resistivity $\rho_0$ (left axis, solid
symbols) and $\rho_0 - \rho_{\mathrm{min}}$ (right axis, open
symbols) as a function of $x$. Insets to (c) and (d) show $T^2$
behavior at low temperatures for the smallest Tl concentrations.}
\end{figure}

The temperature dependence of the normal-state resistivity of the
Tl-doped crystals is shown on an expanded scale in Fig.~\ref{fig2}
for temperatures below 10 K. The resistivity shows a distinct upturn
for temperatures below approximately 9 K, following a form
characteristic of the Kondo effect. Similar results were observed
previously by Andronik and co-workers for temperatures above 4.2 K
for a smaller subset of two Tl concentrations \cite{Andronik_1986}.
A crude estimate for the magnitude of this effect, neglecting any
additional temperature dependence below 10 K, can be made from the
quantity $\rho_0 - \rho_{\mathrm{min}}$, where $\rho_0$ is the
residual resistivity measured at our lowest temperatures and
$\rho_{\mathrm{min}}$ is the value of the resistivity at the
resistance minimum. As shown in the inset to Fig.~\ref{fig2}(a),
this quantity scales approximately linearly with the Tl
concentration $x$. Insets to Fig.~\ref{fig2}(c) and 2(d) show the
resistivity as a function of $T^2$ for the two lowest Tl contents,
$x$ = 0.3 and 0.4$\%$, for which $T_c$ is less than 0.3 K. The
resistivity clearly follows a $T^2$ temperature dependence, as
expected for Kondo-like behavior, for temperatures below
approximately 4 K. In the following paragraphs we argue that this
behavior originates from a charge Kondo effect associated with
degenerate valence states of the Tl impurities. First we demonstrate
that the behavior is not due to localization or electron-electron
effects.

Thallium impurities cause a rapid increase in residual resistivity
of PbTe, characterized by approximately 0.8 m$\Omega\,$cm per
at.$\%$ Tl, as shown in the inset to Fig.~\ref{fig2}(a). Taking $x =
0.4 \%$ as representative, and assuming that the Tl impurities do
not substantially alter the band structure of PbTe \cite{Nimtz,
Khokhlov}, the measured hole concentration of $7 \times 10^{19}$
cm$^{-3}$ implies that there are holes in both the light and heavy
bands located at the $L$ and $\Sigma$ points in the Brillouin zone,
and that the Fermi level lies approximately 180 meV below the top of
the valence band. This allows an estimate of the Fermi velocity,
which has an average value of 10$^6$ m/s for holes in the
anisotropic $L$ pockets and 10$^{5}$ m/s for holes in the heavier
$\Sigma$ pockets. Based on these estimates, the mean free path $l$
is relatively large at 130 \AA{}. With the large resulting values of
$k_{F}l$, at approximately 7 for $x = 0.4 \%$, it is very unlikely
that the low-temperature upturn in resistivity is due to
localization effects. Furthermore, the observed $T^2$ temperature
dependence of the resistivity is not readily identified with such a
scenario.

\begin{figure}
\includegraphics{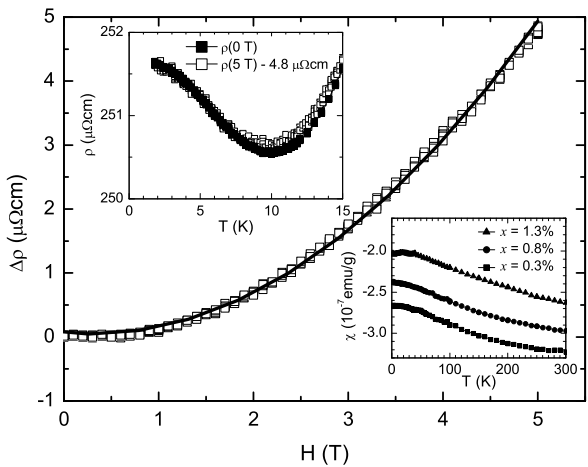}
\caption{\label{fig3}Representative magnetoresistance, $\Delta\rho =
\rho(H)-\rho(H=0)$, for $x = 0.4\%$ at 1.8 K. Upper inset shows
temperature dependence of the resistivity in fields of 0 and 5 T.
The data for 5 T have been shifted down by $\rho_0(\text{5 T}) -
\rho_0(\text{0 T})$ = 4.8 $\mu\Omega\,$cm. Lower inset shows
temperature dependence of susceptibility for different Tl
concentrations. }
\end{figure}

To further probe the origin of the normal-state resistivity anomaly,
the transverse magnetoresistance of the samples was measured, taking
care to avoid spurious Hall contributions. Representative data for
$x = 0.4 \%$ are shown in Fig.~\ref{fig3}. In all cases, the
magnetoresistance is positive, following a $B^2$ dependence for
temperatures above $T_c$ or fields above $H_{c2}$. Furthermore, the
overall temperature dependence of the resistivity shows the same
Kondo-like upturn even in an applied field (upper inset to
Fig.~\ref{fig3}) and is presumably shifted to a higher value due to
a standard Kohler's rule type magnetoresistance. This behavior is
consistent with the absence of both electron-electron effects and
weak localization, which, even in the presence of strong spin-orbit
scattering, would cause a $B^{1/2}$ field dependence at high fields.
Furthermore, this behavior precludes a magnetic Kondo effect as the
origin of the resistivity anomaly, for which a field of 5 T would
cause a substantial negative magnetoresistance.

To verify the absence of magnetic impurities, susceptibility
measurements for several samples of each Tl concentration were made
for arbitrary crystal orientations in an applied field of 1000 Oe
using a commercial Quantum Design SQuID magnetometer. Representative
data are shown in the lower inset to Fig.~\ref{fig3} as a function
of temperature. The susceptibility is diamagnetic for all Tl
concentrations due to the small density of states and becomes less
diamagnetic with increasing hole concentration. The weak temperature
dependence arises from a temperature dependence of both the band gap
and effective mass of PbTe \cite{Lashkarev_1978}. Significantly, the
lack of a Curie-like paramagnetic term from the Tl dopants is
consistent with the absence of magnetic impurities down to $<5$ ppm,
limited by the resolution of the measurement. Hence, the
low-temperature upturn in the resistivity of Tl-doped PbTe follows a
temperature dependence characteristic of the Kondo effect in the
absence of magnetic impurities.

Thallium substitutes on the Pb site in PbTe. Calculations by Weiser
\cite{Weiser_1981} indicate that Tl$^{1+}$ has a lower energy than
Tl$^{3+}$ in the lattice.  Tl impurities therefore initially act as
acceptors, adding one hole per Tl to the valence band, as observed
in Hall measurements \cite{Nemov_1998}. However, the calculated
energy difference between 1+ and 3+ impurity states, which can be
modeled by $\delta E = 2(\epsilon_0 - \mu) + U$ \cite{Dzero_2004}
(where $\epsilon_0$ is the energy to remove an electron from the
6$s$ orbital and $U$ $<$ 0), is very small \cite{Weiser_1981}.
Indeed, for a finite concentration of Tl impurities, the chemical
potential of holes in the system can reach the special value $\mu^*
= \epsilon_0 + U/2$ for which the two valence states become exactly
degenerate ($\delta E = 0$). A value of $\mu$ larger than this would
correspond to all of the impurities being 3+. However, additional Tl
impurities beyond this critical value cannot increase $\mu$ beyond
$\mu^*$ because conversion of all of the impurities to Tl$^{3+}$
would add electrons to the valence band, which would act to reduce
rather than increase $\mu$. Therefore, for Tl concentrations beyond
a characteristic critical value, the chemical potential remains
pinned at the special value $\mu^*$, and any additional Tl
impurities act in a self-compensating manner such that both valence
states are present in equilibrium. This behavior has been confirmed
by Hall measurements, which show that for Tl concentrations beyond
approximately 0.5$\%$ the Hall coefficient saturates to a constant
value corresponding to approximately $10^{20}$ holes per cm$^3$
\cite{Kaidanov_1985, Dzero_2004}. Significantly, the Tl
concentration at which this happens is remarkably close to the
concentration at which we observe the onset of superconductivity
(Fig.~\ref{fig1}). Furthermore, within such a scenario, it is
natural to consider a charge Kondo effect, in which the conduction
electrons interact with the two degenerate valence states of the Tl
impurities, and pseudo-spin flip processes proceed via virtual
excitations to the skipped valence state. In the absence of orbital
degeneracy, the Kondo screening would proceed via a single channel,
so the observation of a resistivity anomaly following a $T^2$
dependence at low temperatures is strong evidence for such a state.

If we associate the observed resistivity upturn of Tl-doped PbTe
with a Kondo-like mechanism, then we can estimate the characteristic
Kondo temperature by fitting the data in the insets to
Fig.~\ref{fig2}c and 2d to
$\rho_{\mathrm{imp}}\sim\rho_{\mathrm{imp}}(0)[1-(\frac{T}{T_{K}})^2]$
where $\rho_{\mathrm{imp}}(0)$ is the impurity contribution to the
resistivity at $T=0$, approximated from the measured values of
$\rho_0 - \rho_{\mathrm{min}}$. This results in a value of $T_K$
$\sim$ 6 K, with considerable uncertainty due to the crude estimate
of $\rho_{\mathrm{imp}}(0)$. Heat capacity measurements involving Na
counterdoping allow an estimate for the range of $\mu^*$ values for
Tl impurities in PbTe, which is characterized by a width of 30 meV
\cite{Nemov_1998}. Assuming a Gaussian distribution of values of
$\mu^*$ centered at 200 meV and with a full width at half maximum of
30 meV, the fraction of Tl impurities for which the two valence
states will be degenerate to within $T_K$ = 6 K is approximately
1$\%$, corresponding to a concentration of $6 \times 10^{17}$
cm$^{-3}$.

From the saturation value of the resistivity we are able to obtain
an estimate of the concentration of Kondo impurities
$c_{\mathrm{imp}}$ using the relation for unitary scattering
\cite{Hewson}, $\rho_{\mathrm{imp}}(0) = 2 m c_{\mathrm{imp}} / (n
e^2 \pi \hbar g(E_F))$, where $n$ is the measured hole concentration
($7\times10^{19}$ cm$^{-3}$ for $x = 0.4\%$) and $g(E_F)$ is the
density of states at the Fermi level (estimated from the band
structure to be 1.4 states/eV/unit cell). We use the effective mass
of the $\Sigma$ band states ($m \sim 0.6 m_0$) since this band
contributes the majority of the density of states. The resulting
estimated concentration of Kondo impurities of $c_{\mathrm{imp}}
\sim 2 \times 10^{17}$ cm$^{-3}$ is consistent with the estimated
concentration of Tl impurities for which the two valence states are
degenerate, within the uncertainty.

In summary, we have measured the resistivity, magnetoresistance, and
susceptibility of single crystals of Tl-doped PbTe,
Pb$_{1-x}$Tl$_x$Te, in the range $0.3 < x < 1.5\%$. We observe an
anomalous low-temperature upturn in the resistivity that scales in
magnitude with the Tl concentration, with a temperature dependence
that is consistent with the Kondo effect. We have demonstrated that
this behavior does not arise from either magnetic impurities or from
localization effects. Given the valence-skipping nature of thallium,
and given that Tl impurities are known to pin the Fermi level in
PbTe, these data are compelling evidence for charge Kondo behavior
associated with degenerate valence states of the impurities. The
effect, which is a unique property of the negative-$U$ Anderson
model, has been predicted theoretically but to date has not been
observed. It appears that PbTe is an ideal host for this effect for
two reasons. Firstly, the special value of the chemical potential
for which the two valence states are degenerate is accessible for
relatively small Tl concentrations. This is likely a consequence of
the requirement of charge balance upon doping Tl impurities into the
largely ionic Pb$^{2+}$Te$^{2-}$ host. Secondly, the large
high-frequency dielectric constant of PbTe, approximately $30-40$
\cite{Nemov_1998}, presumably enables rapid, adiabatic tunneling of
pairs of electrons on and off impurity sites. Most importantly, the
observation of charge Kondo behavior directly attests to an
electronic pairing mechanism for superconductivity in Tl-doped PbTe,
which potentially accounts for the anomalously high $T_c$ value of
this material.

\begin{acknowledgments}

We gratefully thank J. Schmalian, M. Dzero, and B. Moyzhes for
numerous helpful discussions, Robert E. Jones for technical
assistance with EMPA measurements and analysis, N. C. Koshnick, P.
G. Bj\"{o}rnsson, and K. A. Moler for help with dilution fridge
measurements (using an instrument acquired with support from the NSF
under grant no. DMR-9802719), and M. H. Smith for help with sample
preparation. This work is supported by the DOE, Office of Basic
Energy Sciences, under contract number DE-AC03-76SF00515. IRF is
also supported by the Alfred P. Sloan Foundation.
\end{acknowledgments}

\bibliography{Matsushita_condmat_v2}

\end{document}